\documentclass[sigconf,10pt,nonacm]{acmart}
\usepackage{enumitem}
\usepackage{listings}
\usepackage{subfigure}
\usepackage{url}

\setlist[enumerate]{leftmargin=*, parsep=0pt, itemsep=0pt, topsep=2pt}
\setlist[itemize]{leftmargin=*, parsep=0pt, itemsep=0pt, topsep=2pt}

\newcommand{\nej}[1]{{ \textcolor{blue} {Natalie: #1} }}

\newcommand{\jz}[1]{{ \textcolor{purple} {#1} }}

\newcommand{\myparagraph}[1]{\vspace{0.04in}\noindent{\bf {#1}.}}
\newcommand{\mycircle}[1]{\raisebox{.5pt}{\textcircled{\raisebox{-.9pt} {#1}}}}

\title{Accelerator-as-a-Service in Public Clouds: \\An Intra-Host Traffic Management View for Performance Isolation in the Wild}
\author{Jiechen Zhao$^{\varheartsuit}$, Ran Shu$^{\clubsuit}$, Katie Lim$^{\spadesuit}$, Zewen Fan$^{\diamondsuit}$, \\Thomas Anderson$^{\spadesuit}$, Mingyu Gao$^{\diamondsuit}$, Natalie Enright Jerger$^{\varheartsuit}$}
\affiliation{%
  \institution{$^{\varheartsuit}$ University of Toronto, $^{\clubsuit}$ Microsoft Research, $^{\spadesuit}$ University of Washington, $^{\diamondsuit}$ Tsinghua University}%
  \country{}%
}  

\begin{abstract}
I/O devices in public clouds have integrated increasing numbers of hardware accelerators, e.g., AWS Nitro, Azure FPGA and Nvidia BlueField.  
However, such specialized compute (1) is not explicitly accessible to cloud users with performance guarantee, (2) cannot be leveraged simultaneously by both providers and users, unlike general-purpose compute (e.g., CPUs).  
Through ten observations, we present that the fundamental difficulty of democratizing accelerators is insufficient performance isolation support. 
The key obstacles to enforcing accelerator isolation are (1) too many unknown traffic patterns in public clouds and (2) too many possible contention sources in the datapath. 
In this work, instead of scheduling such complex traffic on-the-fly and augmenting isolation support on each system component, we propose to model traffic as network flows and proactively re-shape the traffic to avoid unpredictable contention. 
We discuss the implications of our findings on the design of future I/O management stacks and device interfaces.

\end{abstract}

\begin{document}

\maketitle

\section{Introduction}

%\nej{In 1-2 sentences in as straightforward a manner as possible, what is the problem that is being identified in this paper?}

%\nej{With shorter papers, you do need to get to the point faster than in a longer paper. However, I find the intro as written makes it very difficult to understand what is the problem that needs to be addressed and why is it important. The intro gets into the weeds very quickly without framing the bigger picture for the reader.}

To serve the growing demand for online software services~\cite{tail-at-scale}, CPU efficiency is increasingly crucial due to stagnating performance~\cite{esmaeilzadeh2011darksilicon}. 
Cloud providers have found that some common tasks, such as encryption, compression, and hashing, can consume up to 82\% CPU cycles~\cite{kanev2015profiling,sriraman2020accelerometer,karandikar2023cdpu,karandikar2021grpcacceleration}. 
To save CPU cycles and improve overall application throughput, clouds have started to leverage \textit{hardware acceleration}~\cite{karandikar2023cdpu,wolnikowski2021zerializer,kwon2020fvm,azuresmartnic,aws-aqua,taylor2020asiccloud} to offload those auxiliary tasks.  
Fortunately, a diversity of bespoke silicon units already exist in PCIe-attached I/O devices for providers to use~\cite{dpu-3,amd-xilinx-smartnic,galles2021pensando,intel-ipu,aws-nitro,intel-quickassist}. 
%I/O devices with accelerators can be those in Table~\ref{table:accelerator-contention-example}, or others like AWS Nitro and Azure FPGA. 

Meanwhile, \textit{users} of public clouds also suffer from a similar low-efficiency problem. 
%While accelerators improve performance for providers' operations, their benefits have not been delivered to \textit{users}
%\nej{needs rewording for grammar as well as clarity. Also, personally, I don't like the use of "on the other hand" if "on one hand" hasn't appeared first}
For example, users need their own cryptography operations for confidentiality~\cite{azure-confidential-vm}, which slows down up to 4,000$\times$ if CPUs rather than accelerators execute the homomorphic encryption algorithm~\cite{samardzic2022craterlake}.
%without  this adds 4,000$\times$ overheads to users if they prefer no data leakage when exchanging data with an untrusted server. 
%\jz{Add why tenants and providers want this: heterogeneous computing resources are users' need. For examples, service provider VM nested, unconfidential. Web service in cloud needs Firewall (hardware is more feasible than writing their own). }
%\nej{"For example" sentence and "Although" sentence are not well-connected.  They seem to be about different topics. Also, first paragraph introduced the existence of such accelerators.}
%Although such accelerator resources already existed \kl{exist where? on smartNICs?} in the public cloud, 
However, existing cloud accelerators accessible to \textit{users} do not support performance guarantees~\cite{aws-f1}. %\jz{Explain more. Use a table like \ref{table:providers}}
In addition, some accelerators can be needed by both \textit{users} and \textit{providers}.
Unfortunately, %due to poor manageability, 
they currently cannot be leveraged simultaneously by the two parties, unlike general-purpose compute (e.g., CPUs). %\jz{Tasks are similar leads to the goal of using them simultaneously. }
%\nej{can you reword part 2? "have not been discussed" -- can this be: currently cannot be leveraged simulatenously...}

%\input{tables/provider-accelerators}

%% ***** writing reminders *****
%\begin{itemize}
%    \item Mention accelerators already exist in the cloud [done]. 
%    \item Show a typical system architecture: GPU, CPU, SSD, NIC, accelerator
%    \item Clarify that both VM users and provider can be benefited from accelerators. Just like CPUs. 
%\end{itemize}

\myparagraph{Design goals}
In this paper, 
we aim to provide accelerators on diverse devices \textit{as a service} to the users in the public cloud, allowing users to specify requirements in the same way they can specify how many CPU cores to rent. 
Our proposal  addresses
\mycircle{1} how accelerators can be used under a \textit{diversity} of scenarios (network, storage, security, etc.) on various I/O paths and 
\mycircle{2} how providers and multiple users can \textit{simultaneously} use each accelerator (when available) with \textit{end-to-end performance isolation}. 
%\nej{I'm not sure that address is the best verb -- this is more of a characterization/observation and open questions paper so it at present doesn't present a solution that addresses this issue.} 
%\rs{I don't think we "emphasize" this, we can just say it as a fact. The fact should belongs to the previous paragraph}

%We strive to achieve performance isolation in such a complex usage model and tackle the challenges which will be discussed next. \nej{I don't feel that this sentence adds much. I'm not clear on the "complex usage model" that you refer to. I would consider removing this sentence or focus on the need for performance isolation.}
%\nej{I agree to define the two modes}
%\nej{this is generally pretty clear (I agree it would be good to define the two modes). I find most of the intro prior to this to be difficult to wade through. I would consider starting with the intro at a higher level and identifying your goal prior to diving into a lot of details.}

\myparagraph{Relevant techniques}
%\kl{It's a bit unclear why we're focusing so hard on NICs at this point}\jz{I rephrased some sentences. }
Although no current proposals support both design goals, %\nej{"both goals" is a bit unclear -- what are the two goals laid out in the previous paragraph} 
there are some techniques that support multi-tenancy I/Os for bandwidth allocation, and some techniques for leveraging I/O accelerators on NICs or storage devices. 
%\jz{No one can achieve this goal, but some techniques have some relations with multi-tenancy.} 
%\jz{Mention that some technqiues are not designed for public clouds (limited knowledge). Some are not designed for accelerators. Some are not designed as services. }
Unfortunately, some multi-tenancy proposals for I/O bandwidth need detailed tenant request information~\cite{zhang2022justitia,kong2023understanding-rdma-microarch,min2021gimbal,khalilov2023osmosis}, which is not available in the public cloud.
Some fail to satisfy diverse scenarios because they tightly couple accelerators within only network or storage I/O paths and consider only simple traffic patterns~\cite{lin2020panic,liu2019ipipe,grant2020fairnic,kwon2021flexcsv,eran2022flexdriver,eran2019nica,ruan2019insider}. 
%Some emphasize IOMMU and memory bandwidth related congestion~\cite{agarwal2022understandinghostinterconnect}, %\nej{I don't really know what you mean by demystify IOMMU and memory bandwidth contention?} 
%but not other components like device interface and PCIe interconnects on the datapath. 
Others allow accelerator sharing but fail to offer service-level agreements (SLAs) ~\cite{lin2020panic,liu2019ipipe,grant2020fairnic}.
Finally, no proposals take the characteristics of accelerators themselves into account.

\myparagraph{Understanding isolation breakage}
%\jz{Research find that existing accelerators in the cloud, e.g., AWS F1 FPGAs, is likely to co-locate tenants on the same device~\cite{tian2021cloudfpga-pcie-contention}. }
Table~\ref{table:accelerator-contention-example} shows the end-to-end accelerator throughput ratios on three devices. 
In each test, two co-located tenants with the same priority invoke an accelerator under various traffic patterns. %with the same priority. 
%Both tenants invoke the same I/O accelerator in function call mode. 
Instead of equal allocation (i.e., ratio=1), we find performance variability widely exists on different accelerators. 
%\nej{Not really clear on what the table is telling me. What is the case study that is evaluated?} 
%\mg{Also, it seems the table now shows three apps (acc. type?) on three devices, not ``two co-located tenants invoking the same accelerator'' as said above. Unclear to me.}\jz{We run two co-located tenants to contend one accelerator (this is one run). Then we test multiple runs on different devices. i see the clarity issue you mention and I'll work on describing it in a better way }

The fundamental challenges for performance isolation are twofold. 
%First, traffic patterns are unknown a priori in the public cloud, and they are too diverse to manage. For example, a combined pattern\footnote{} from two unknown tenants' traffic streams is mixed and hard to predict. % message rates and sizes are different in network and storage domains.  
First, traffic patterns are too diverse to manage. 
Even though an individual tenant is unlikely to be problematic, unpredictable contention still can happen depending on what traffic it is co-located with\footnote{Unless specified, the ``traffic pattern" in this paper refers to the mixed pattern of all tenants rather than that of an individual tenant's traffic. }. 
Worse, such mixed patterns are unknown a priori in the public cloud.
Second, multiple possible contention sources on I/O datapath\footnote{System components on those paths include: \mycircle{1} host CPU, \mycircle{2} host network (e.g., root complex, PCIe interconnects), \mycircle{3} accelerator-side interface (e.g., buffers, queues, schedulers), and \mycircle{4} heterogeneous accelerators themselves. } can become the bottleneck and lead to isolation breakage. 
Accelerator I/O operations can also contend with other legacy I/O operations such as those from network and storage. 
Removing all possible contention sources may be possible but would be complex.

In Sec.~\ref{sec:understanding}, we break down potential contention sources in diverse I/O paths and then analyze them under various multi-tenant traffic patterns. 
Based on our observations, the root causes of the results in Table~\ref{table:accelerator-contention-example} are \mycircle{1} the limited low-level isolation mechanisms of components %(e.g., PCIe interconnects, device interfaces)
on those paths (e.g., tenants' traffic not isolated across PCIe lanes but allocated by credits), and \mycircle{2} not taking computational heterogeneity and non-linearity\footnote{Non-linearity means accelerators I/O operations are not ``channels" like read/write I/O or send/receive I/O that exhibit linearity, e.g., equal egress/ingress bandwidth requirements, and larger bandwidth can be linearly allocated for larger messages~\cite{grant2020fairnic,lin2020panic,eran2019nica,min2021gimbal,liu2019ipipe}. } %\kl{i don't think we've defined linearity yet}
of accelerators into account.

\begin{table}
    \centering
    \small
    \caption{Case studies: unpredictable accelerator throughput allocation ratios.%\nej{bluefield row is missing a number?}
    }
    %\vspace{-1em}
    \begin{tabular}{ccc} 
    \toprule
    \textbf{Device} & \textbf{Acc. Type} & \textbf{Thr. Ratio} \\
    \midrule
    Intel QuickAssist ASIC & Compression &  0.5-5.1$\times$  \\ 
    Nvidia BlueField-2 & RegEx &  0.31-6.5$\times$  \\
    Xilinx Alveo U250 FPGA & Hashing &  0.25-8.4$\times$  \\
    \bottomrule
    \end{tabular}
    \label{table:accelerator-contention-example}
\end{table}

\myparagraph{Our proposal: traffic shaping}
%\nej{similar to above -- this is generally clear -- can we get to this without getting to into the weeds on the "fundamental challenges" that you outline above? And move that material on fundamental challenges elsewhere?}\jz{I've removed the summarized challenges and put this paragarph below a new paragraph, which is the observed isolation breakage for several types of accelerators.}
Instead of improving per-component isolation mechanisms by re-designing CPU boards and devices, we treat accelerator-related I/O as \textit{traffic flows} and perform \textit{accurate traffic shaping}. 
%This is motivated and guided by our characterizations of possible contention sources within I/O paths as well as accelerator heterogeneity (Sec.~\ref{sec:understanding}). 
This idea is inspired by the effectiveness of per-NIC traffic shaping in datacenter networks for isolation enforcement and traffic management~\cite{radhakrishnan2014senic,ballani2011oktopus,kumar2019picnic,angel2014pulsar}. 
%\nej{"by the significance" is vague -- what specifically are you referring to here?}
We consider the untrusted accelerator I/O flows transferring within a server through insufficiently isolated infrastructure as similar to the datacenter network setting.   
Therefore, our traffic shaping approach is well-suited to addressing multi-tenancy isolation problems within a heterogeneous server architecture. 
We propose a system-level accelerator management stack that offers a decoupled management interface for accelerators. 
We discuss the feasibility of where and how to perform traffic shaping (Sec.~\ref{sec:design}), as well as other open problems (Sec.~\ref{sec:open}). % in terms of permission authentications, isolation manageability, SLA accuracy, and applicability for large-scale deployments. \jz{Sec5}

\begin{figure}
  \centering
%  \hspace{-3em}
  \includegraphics[width=.99\columnwidth]{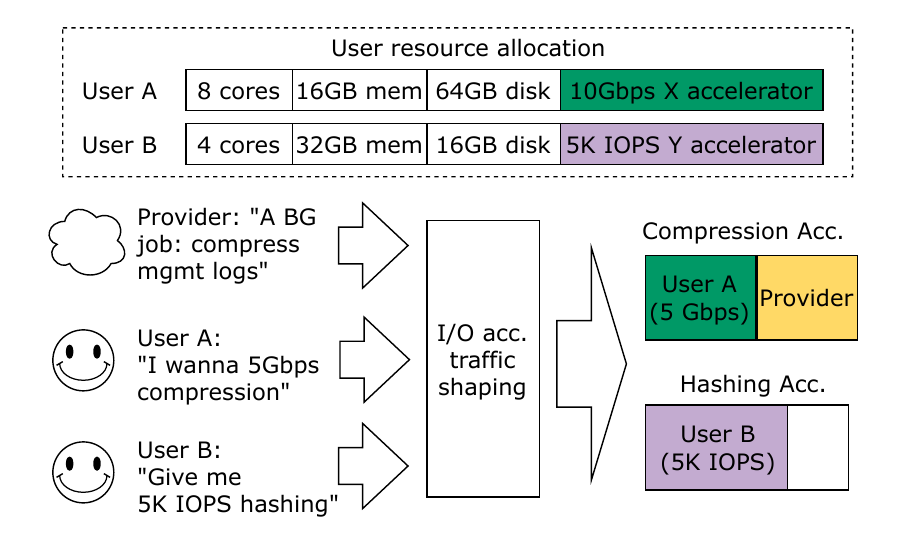}
  %\vspace{-1.5em}
  \caption{New services empowered by our proposal.
  }
  %\vspace{-1.5em}
  \label{fig:service}
\end{figure}

\myparagraph{Acceleration service enabled by our proposal} %\nej{rather than vaguely say "new feature" -- I would concisely describe or name the feature}
%This proposal can enable a new feature for the public cloud. \nej{this first sentence is redundant with the paragraph heading}
Fig.~\ref{fig:service} shows how this feature works. 
Users can directly request the type and throughput of accelerators, just like the existing ability to request the number of cores.
For a given accelerator, we allow multiple users and the provider to take full advantage of it, %\mg{missing words?} 
without worrying about performance isolation breakage (e.g., compression accelerator in Fig.~\ref{fig:service}).
Because our traffic shaping approach takes flow characteristics into account (e.g., message sizes and rates), we enable users to request either based on Gbps or IOPS (similar as the storage performance business model in today's clouds~\cite{azure-storage-disktypes}). 
Thus, users in multiple scenarios can safely invoke this service. 

Note that our paper targets the infrastructure-as-a-service (IaaS) model for accelerators.
We assume the cloud provider is trusted.
All accelerators are offered and managed by the provider. 
Tenant users cannot program the accelerators.
VM users are untrusted in terms of how much accelerator traffic they invoke.

\section{Background}\label{sec:background}

\begin{figure*}[t]
\centering 
\subfigure[AES-256-CTR]{
    \includegraphics[width=0.17\textwidth]{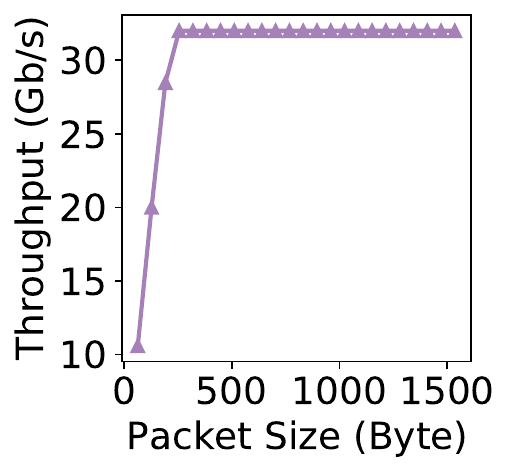}}
\subfigure[IPSec]{
    \includegraphics[width=0.17\textwidth]{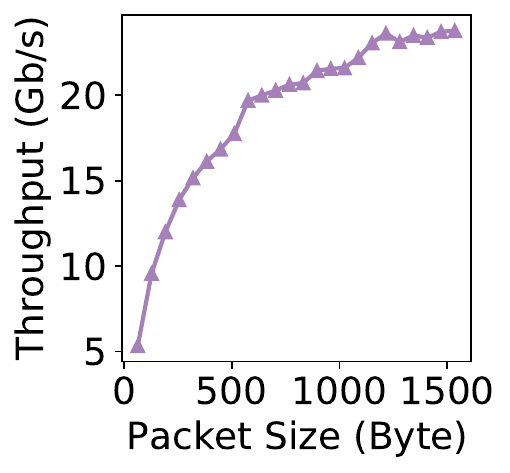}
}
\subfigure[SHA-3-512]{
    \includegraphics[width=0.17\textwidth]{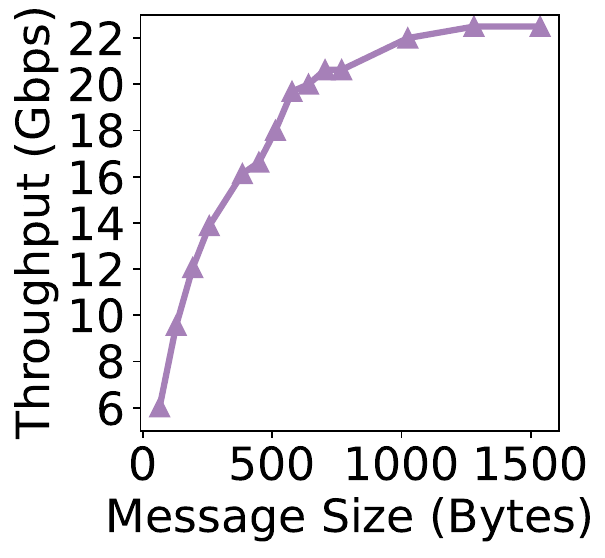}
}
\subfigure[Regex]{
    \includegraphics[width=0.17\textwidth]{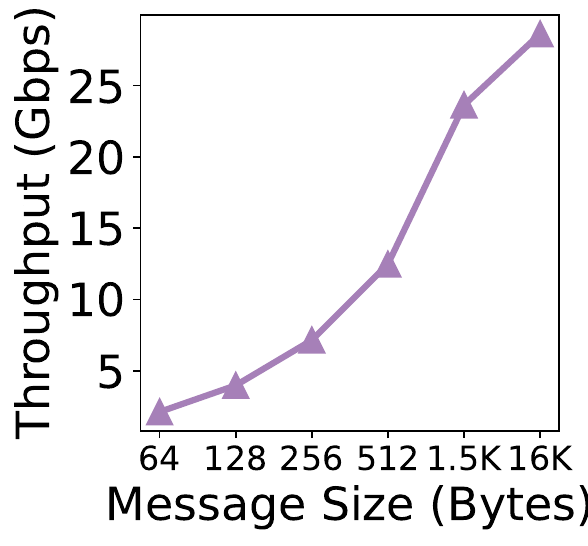}
}
\subfigure[QATzip]{
    \includegraphics[width=0.17\textwidth]{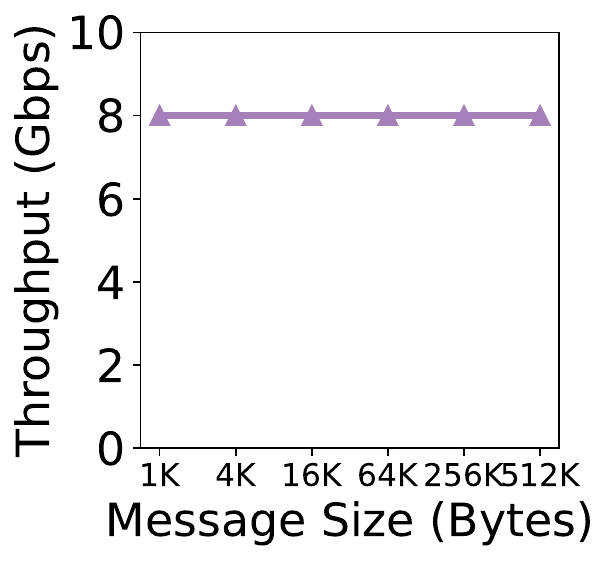}
}
%\vspace{-1.1em}
\caption{Accelerator profiling: compute throughput across message sizes.
}
\label{fig:profiling}
\end{figure*}

\myparagraph{Benefits of I/O accelerators in public clouds}
Existing cloud infrastructures are embracing supports for accelerators, such as AWS Nitro~\cite{aws-nitro}, Azure FPGA~\cite{azuresmartnic} and Alibaba Cloud CIPU~\cite{alibaba-cipu}. %\jz{Check IBM and Oracle}
%\footnote{GPUs or TPUs are not in the scope of I/O accelerators in this paper.}
This trend is driven by the higher overall throughput and improved system efficiency benefited from accelerators. 
For instance, it is possible to offload 3-15\% of CPU cycles by using (de)compression accelerators~\cite{karandikar2023cdpu}.
Other work has shown that an IPSec accelerator offering 32Gbps throughput~\cite{lin2020panic} dramatically outperforms CPUs, which can only deliver 10Gbps while consuming as many as 8 Xeon cores.
In addition, a homomorphic encryption accelerator allows a user to exchange data with an untrusted server 4,000$\times$ faster than a CPU~\cite{samardzic2022craterlake}. 

In the future, we expect accelerators to be even more prevalent. 
Their growing importance can be seen from the increasing numbers and types of accelerators from the first to the third generation Nvidia BlueField DPUs~\cite{dpu-3}.
%\nej{how does this show the growing importance?}\jz{I would add more evidence at this point}
Similarly, the Pensando Elba chip devotes significant die area for specialized processing engines and components~\cite{Pensando-floorplan}. 

\begin{comment}
In most parts of this paper, we take accelerators on SmartNICs as a key example of I/O accelerators. 
SmartNICs have become a key part of cloud infrastructures~\cite{aws-nitro,azuresmartnic,alibaba-cipu}, and accelerators have been off-the-shelf in many SmartNIC products~\cite{dpu-3,amd-xilinx-smartnic,galles2021pensando,intel-ipu} and SmartNIC infrastructure customized by cloud vendors~\cite{azuresmartnic}.
GPUs or TPUs are not in the scope of I/O accelerators in this paper.
\end{comment}

%\footnote{Accelerators such as GPUs or TPUs are not the focus of this paper.}

%%%% ****** writing reminder: acc opportunity table ***** 
%\input{./tables/smartnic-acceleration-opportunity}

%\subsection{Host Protocols for Diverse I/O Accelerator Paths}

\myparagraph{CPU-accelerator protocol}
The most commonly used protocol between the host and high-performance devices (e.g., NICs, SSDs, and accelerators) is based on the ring buffer abstraction, a \textit{producer-consumer protocol}~\cite{neugebauer2018understandingpcie,kwon2020fvm,dpdk-compression-accelerator,dpdk-crypto-accelerator}. 
%\nej{this should have some citation to corroborate the "most commonly used" claim}
The device and host software communicate by exchanging descriptors.
% Omit some information below: 
%Descriptors hold metadata, including packet size, what processing the NIC should perform (e.g., update the checksum or segment the packet), a fag bit, and a pointer to a separate packet buffer which holds the actual packet data. 
On the device side, multiple hardware I/O queue pairs (QPs) are used to maximize bandwidth utilization. 
Each QP (submission queues (SQs) and completion queues (CQs)) is associated with a set of doorbell registers addressable via host-side load/store instructions and mapped to the physical memory address space. % omit the concept of PCIe BAR regions and MMIO
The host side pre-allocates DMA buffers and descriptor ring buffers for each hardware queue.
% omit information for host-side maximizing parallelism - each QP on each core; each buffer slot size is fixed; 

This process involves: \mycircle{1} The host sends a notification to ring the doorbell register of a particular hardware queue. Then \mycircle{2} the device uses DMA to read pre-prepared descriptors from the corresponding SQ ring buffer, \mycircle{3} reads data for the accelerator from the appropriate DMA buffer, and finally, \mycircle{4} updates the CQ ring buffer pointer using a DMA write.  
%We refer to such an accelerator invocation as \textit{accelerator I/O}. 

%\jz{Consider if add a figure}

%\myparagraph{Network-accelerator protocol}
%\jz{Only put some points here. Haven't rephrased them. }
%There is no standard way to conduct this protocol. 
%Network-attached accelerators on SmartNICs have been widely studied~\cite{beehive}. 
%SmartNIC accelerators are originally designed to accelerate network applications~\cite{}. 
%They use either DMA between the network logic and the accelerators. 
%Here in this paper, we are the first to consider that SmartNIC accelerators do not necessarily work for network applications in an inline mode, but can work for multiple application domains or infrastructure tasks. 

%\myparagraph{Storage-accelerator protocol}
%\jz{Only put some points here. Haven't rephrased them. }
%Storage accelerators integrated in the SSD~\cite{atc}. 
%In SmartNIC, SmartNIC accelerators can directly communicate with the SSD if the accelerator logics can interpret NVMe commands and payload and header. 

\section{Embracing Accelerators: Isolation Problems in the Wild}\label{sec:understanding}

This section understands the performance isolation problem for accelerators. 
After describing the methodologies (Sec.~\ref{characterization:method}), this paper studies the components involved on accelerator paths and analyze the results regarding computation (Sec.~\ref{characterization:accelerators}) and communications (Sec.~\ref{characterization:commoditynic}). 
Further breakdown analysis are given in Sec.~\ref{characterization:fpga}. 

%\jz{better connections and better descriptions on what the flows are. }

\subsection{Methodology}\label{characterization:method}

Since available commercial devices lack of sufficient visibility for multi-tenancy study, we set up hybrid testbeds.
We separately study the characteristics of components along I/O paths, including accelerators (Sec.~\ref{characterization:accelerators}), and the host network (e.g., root complex, PCIe interconnects) and device-side interface (e.g., hardware QPs, on-device buffers) (Sec.~\ref{characterization:commoditynic}-\ref{characterization:fpga}).

\myparagraph{Accelerators} 
We build a sub-system on an Intel Arria 10 FPGA. 
A traffic generator feeds messages into a 256KB data buffer unless the buffer is full. 
The accelerator pulls the next request from the data buffer before the current computation finishes. The pulling is in FIFO order. These results are analyzed in Sec.~\ref{characterization:accelerators}. %% \jz{Add accelerator descriptions or summarize them in a table}
%Due to the performance bottleneck in the kernel-based FPGA driver~\cite{forencich2020corundum}, we use hardware counters on the FPGA to report accelerator throughput instead of capturing messages on the host machine.
%\mg{This description seems to offer no info about the accelerator logic itself, like functionality and throughput. Maybe at least give its throughput?}

% The accelerator introduction one by one
\begin{comment}
    
\textit{IPSec}. IPSec helps protect the confidentiality and integrity of information as it travels across less-trusted networks in the public cloud. 

\textit{AES}. The AES algorithm is a symmetric block cipher that can encrypt (encipher) and decrypt (decipher) information. Encryption converts data to an unintelligible form as ciphertext; decrypting the ciphertext converts the data back into its original form, i.e., plaintext.

\textit{SHA}. SHA is one of the complex hash functions that are widely used. It is a ``one-way" cryptographic function and is a fixed size for any size of source text. This makes it suitable when it is appropriate to compare ‘hashed’ versions of texts, as opposed to decrypting the text to obtain the original version.

\end{comment}

%\nej{The previous section already presented some data without providing a methodology}
\myparagraph{Commercial devices}
Besides accelerators' computational characteristics, we also need to study the contention impacts of certain traffic patterns on the device interface and the host networks. 
We choose one commodity NIC for characterizations due to the following reasons. 
First, host-NIC systems have already established mature software and hardware support for optimized host-device communications. 
Second, this testbed exhbits representative datapaths for accelerator invocations. 
This is because leveraging the NIC corroborates with prior work which reuses the NIC interface for accelerator invocations ~\cite{eran2022flexdriver,eran2019nica,lin2020panic,grant2020fairnic,khalilov2023osmosis}. 
%\nej{You need another sentence here -- the reader doesn't know what this section is going to be about}
%We use the following system setting to study contention issues within the host network plus device interface. 

Specifically, the host has dual 16-core Intel Xeon E5 2698v3 CPU sockets and 256GB RAM, running Ubuntu 18.04 with Linux kernel version 4.17.12 installed. 
%No DDIO enabled. 
We attach one Mellanox ConnectX-6 card to this server and use its one 100Gbps port. 
To prevent the contention impacts from host resource contention (e.g., shared caches, memory controllers, and memory channels) from affecting our contention study, we run each tenant on a separate NUMA node. 
%The IOMMU is enabled. 

%\jz{We get rid of the contention effect between CPU-memory traffic and PCIe-memory traffic. This work~\cite{agarwal2022understandinghostinterconnect} finds memory bus are shared thus suffering from contention even for single tenant.}
%\nej{I'm unclear on the goal of this section -- what are you intending to convey}
%\jz{I reorg the first two paragraphs a bit}
%% writing reminder: Detailed mothodology see the Google Sheet 

Each tenant sends arbitrary sizes of intra-host RDMA reads or RDMA writes, which invoke host-device round trips based on the ring buffer protocol described in Sec.~\ref{sec:background}. 
Writes and reads of RDMA represent individual traffic in this round trip, i.e., device-to-host heavy and host-to-device heavy respectively. 
We set the same priority for both tenants. 
Each test runs for 10 minutes, and observations are in Sec.~\ref{characterization:commoditynic}. 
%This mimics the DMA reads and writes within the server. We denote workloads as DMA. 
%\jz{You should name this as round trips. }

\myparagraph{Host-FPGA prototype}
%The software of commodity devices like drivers and firmware is tightly coupled with device vendors' implementation. 
%The hardware of commodity devices and host networks are ``black-boxes" to the provider that manages I/O multi-tenancy\jz{Is that a valid statement? }. 
Given the difficulties of changing low-level details within commercial devices, we create a host-FPGA prototype to further break down our analysis in Sec.~\ref{characterization:fpga}. 
By customizing the driver on the host and the FPGA as the I/O device, we build an end-to-end system with better observability on each component from tenant processes running on the host to an accelerator. 
Moreover, we can mitigate some factors in the prototype to give further insights for device vendors (as accelerator designers) and cloud providers (as accelerator managers).\footnote{We expect that there will be a hybrid model where providers may customize their devices, and meanwhile they buy third-party devices from vendors.} 
We use an Intel Arria 10 FPGA connected to the CPU host above via a PCIe root complex with a PCIe Gen 3.0 x8 interface. 
Each test runs 1 million commands and results are presented in Sec.~\ref{characterization:fpga}. 
%\mg{This whole paragraph seems to be redundant with the previous paragraphs in this section, e.g., difficulty of looking into commercial devices, the Arria 10 FPGA, the separate analysis of each component alone.}

\subsection{Computational Characteristics}\label{characterization:accelerators}

We emphasize two %\mg{which ``three'' are you talking about? This subsection has 2, and in total there are 9.} 
\textbf{features that have been overlooked} by previous studies on multi-tenant accelerators.

\myparagraph{Observation 1: Message size matters}
%This paper emphasizes the importance of understanding performance trends with message sizes fed into each type of accelerator. 
Fig.~\ref{fig:profiling} shows the unique (non-linear) relations between each accelerator's compute throughput and its input data sizes. 
We point out that accelerators usually cannot deliver throughput linear to the data granularity they are fed, unlike network link bandwidth allocation.\footnote{This assumption is naturally true for send/receive I/Os and read/write I/Os, though not explicitly clarified by prior multi-tenancy I/O work.} 
Instead, such a trend is ad-hoc for each accelerator type. 

%This feature is overlooked by prior multi-tenant I/O sharing proposals~\cite{grant2020fairnic,lin2020panic,eran2019nica,tork2020lynx,liu2019ipipe} because they still treat accelerator invocations as ``packets". 
%Thus, some commonly used packet scheduling policies, such as Weighted Fair Queuing (WFQ) and Deficit Round Robin (DRR), are not directly applicable to precise allocation and performance isolation for heterogeneous I/O accelerators. 
%\nej{can you explain this further?}

\myparagraph{Observation 2: Egress/ingress bandwidth ratio varies}
We highlight that the bandwidth requirements of an accelerator invocation can differ in terms of ingress and egress paths. 
Based on our observations, there are several possible value ranges for the ratio $\frac{egressbw}{ingressbw}$ (denoted as $R$=$\frac{Eb}{Ib}$).  

\begin{itemize}
    \item $R$=1. For example, the output ciphertext of AES-256-CTR is always the same length as the input plaintext. 
    \item $R$>1, e.g., decompression falls into this category. 
    \item $R$<1, e.g., compression falls into this category.  
    \item $Eb$ is fixed. For example, SHA-3-512 has a fixed output message size of 64B, no matter how large the input is. 
\end{itemize}

These categories lead to different potential contention cases. For examples, SHA-3-512 is much more likely to interfere with its ingress path (i.e., DMA reads) rather than on its egress path (i.e., DMA writes), which always has small messages as outputs; 
allocating $X$ Gbps PCIe bandwidth might not be sufficient to feed data into a compression accelerator where a user requires $X$Gbps. 
%\nej{I would rewrite the example to be more clear -- the output message is always small which is why it is less likely to see contention than the input?}

%% Writing reminder: omit implemnetation details

\myparagraph{Observation 3: Implementation matters}
Even for the same type of accelerator, different algorithms (e.g., lossy vs. lossless compression) will change the impact of the above curves. 
Additionally, different hardware implementations (e.g., %Verilog implementations, \nej{not sure the reason to mention Verilog implementation here -- that Verilog would be realized as either an ASIC or on an FPGA}, 
ASIC vs.~FPGA prototypes) will also impact the curves. 
The observation indicates that the providers should be aware of the characteristics of the accelerators in use. 
There should also be a rethinking about users' observability: users need to have a nicely designed management interface to know details about the accelerator they pay for. 

%\jz{I got examples comparing FPGA and Intel QuickAssist ASIC accelerators.}
%\nej{I don't find it particularly surprising that different implementations will have different characteristics -- is there something surprising or some new insight that I am missing here?}
%\jz{Yes, I would add the challenge that the implementation difference exists, so the users cannot easily get the characteristics of accelerators that they are using, difference (or they don't have the permission to know), the providers also cannot naively treat all accelerators with the same traffic shaping policy even though the accelerators are all the same type. }

\subsection{Commercial Device Characteristics}\label{characterization:commoditynic}

We observe that the host network and the device interface collaboratively affect end-to-end isolation. 

%\jz{TODO: summarize data in IOMMU disable case}

%we use RDMA that bypasses the CPU. 
%Each RDMA tenant runs on a separate NUMA node to further isolate the contention of the two tenants on shared caches, memory controllers, and memory channels. 

%% Writing reminder: This full figure is trimmed down to the following single column figure mainly for space
\begin{comment}

\begin{figure*}
\centering 
\subfigure{
    \includegraphics[width=0.4\textwidth]{./figures/qp-num1.pdf}
}
\subfigure{
    \includegraphics[width=0.32\textwidth]{./figures/qp-num2.pdf}
}
\subfigure{
    \includegraphics[width=0.24\textwidth]{./figures/qp-num4.pdf}
}
\vspace{-2
em}
\caption{QP number study on contention. Tenant A (4KB) V.S. Tenant B (4KB)
\nej{is this result unexpected? what is the ideal or expected result? What is interesting about this?}}
\jz{TODO: Reduce the number of sub-graphs. Annotate each sub graph like Fig.~\ref{fig:msg-size-4KB}. Change it to be X QP: X QP, 2X QP: X QP, 4X QP: X QP, 8X QP: X QP, 16X QP: X QP}
\label{fig:qp-num}
\end{figure*}
\end{comment}

\myparagraph{Observation 4: Different QP numbers impact bandwidth allocation} 
We sweep the QP number from 1 to 16 for each tenant.
Based on our explorations on two tenants both sending 4KB RDMA writes, we find the throughput ratio between two tenants ($T_{q}$) is equal to the ratio of QP numbers.
For example, 2QP vs.~1QP, and 8QP vs.~4QP exhibit the same bandwidth allocation result, i.e., $\sim$52Gbps vs.~$\sim$26Gbps. 
Other cases are similarly linear but data is omitted. 
A straightforward estimation is that the round robin arbitration commonly used by commercial devices~\cite{intel-quickassist,li2019prioritygpu,martinasso2016ethpciecongestionongpus} leads to this QP-related linear throughput allocation. 
This observation suggests the provider must be aware of per-tenant QP allocation and be able to re-allocate QP numbers.\footnote{Previous multi-tenant RDMA NICs do not consider QP impacts and their reallocation~\cite{zhang2022justitia,kong2023understanding-rdma-microarch}.}

\myparagraph{Observation 5: Different message size mixtures contend differently}
In this test, both tenants have one QP allocated on the NIC and both send RDMA writes.
Tenant A sends 4KB messages all the time, while Tenant B varies its message sizes from 256B to 8KB. 
We find the throughput ratio $T_{m}$ varies from 0.83$\times$ to 10.5$\times$, exhibiting throughput allocation ratios that are non-linear 
to message size ratios. 
This may be due to (1) NIC interface contention (e.g., on buffers or caches), and (2) fair queuing policy giving more PCIe bandwidth to larger messages.

\myparagraph{Observation 6: Direction of traffic heaviness matters}
The heaviness of data transfers has directions, i.e., host to accelerator (HtA), or accelerator to host (AtH). 
We study the contention effects of two scenarios: (1) RDMA writes colocate with RDMA writes (i.e., homogeneous heaviness), and (2) RDMA writes colocate with RDMA reads (i.e., heterogeneous heaviness). 
%We set the tenants with the same priority and expect them to get the same end-to-end throughput. 
All experiments use 2 QPs and the same message sizes (from 256B to 8KB) for both tenants. 
For homogeneous heaviness, the bandwidth is evenly allocated. %, satisfying our expectation.%\footnote{The throughput differences in 512B and 256B cases are due to NUMA effects (Tenant A on NUMA0 with the NIC attached, Tenant B on NUMA1).} 
However, for heterogeneous heaviness, AtH traffic steals 1.08-3.39$\times$ bandwidth from HtA channels.
Thus, the accelerator management stack should be aware of the direction of heavy traffic.

\subsection{Breakdown Analysis}
\label{characterization:fpga}

\begin{figure*}
\centering 
\subfigure[IOPS vs. Gbps isolation (both DMA reads)]{
    \includegraphics[width=0.38\textwidth]{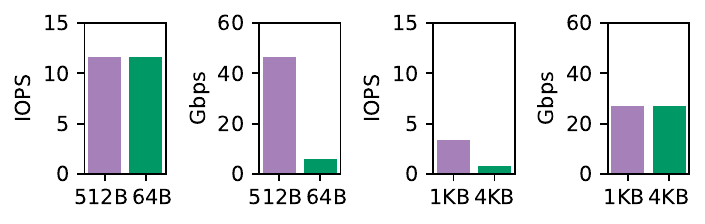}}
\subfigure[Stable read-write mixture]{
    \includegraphics[width=0.18\textwidth]{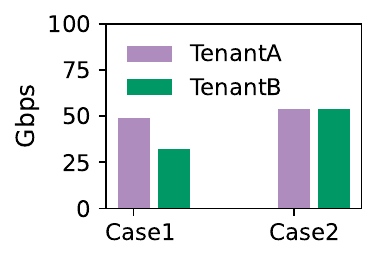}
}
\subfigure[Harmful tiny messages]{
    \includegraphics[width=0.18\textwidth]{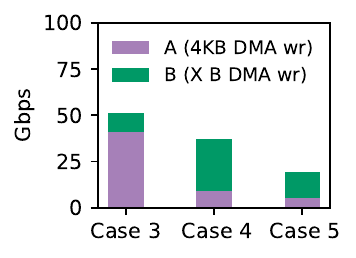}
}
\hspace{-1em}
\subfigure[Harmful tiny messages]{
    \includegraphics[width=0.18\textwidth]{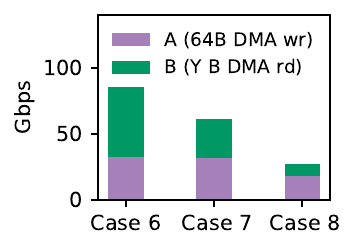}
}
%\vspace{-1em}
\caption{Host-FPGA characterization results. 
%\vspace{-1em}
}
\label{fig:fpga-results}
\end{figure*}

Within our host-FPGA prototype, we list the following features to better understand the phenomena in Sec.~\ref{characterization:commoditynic}. 
\begin{itemize}
    \item \textbf{Ring buffer protocol breakdown}. We manually implement PCIe read and PCIe write commands, supported by an FPGA driver and a register interface on the FPGA. This allows us to break down the complex communication in the ring buffer protocol that requires multiple round trips in the host network. The breakdown helps contention source analysis in detail. 
    \item \textbf{Awareness on arbitrary on message sizes and PCIe MTU size}. 
    %\mg{``Arbitrary awareness'' sounds a weird term. Maybe ``Awareness on arbitrary message sizes ...''?}
    To tune accelerator message sizes injected into the host network, we build a DMA engine on the FPGA that can split accelerator messages into arbitrary sizes of PCIe reads and writes. The engine is also aware of PCIe ``MTU" size\footnote{We refer  $MaxReadReqSize$ and $MaxPayloadSize$ to be PCIe ``MTU" similar to Ethernet MTU. } when splitting accelerator DMA messages. 
    \item \textbf{Arbitrary QP management}. We manually implement hardware QPs and study contention effects on different arbitration or fair queuing policies across tenants.  %manually allocate host memory regions for each tenant to invoke an accelerator on the FPGA.   
    \item \textbf{End-to-end invocations for each tenant}. We run two tenants on the host that interact with the driver. Reported throughput is end-to-end, from tenant processes instead of only from FPGA hardware counters~\cite{lin2020panic}.
\end{itemize}

%% writing reminder: Detailed mothodology see the Google Sheet 

%Intuitively, a straightforward estimation is that the robin robin arbitration commonly used by commercial devices~\cite{mellanox,intel-quickassist} leads to the problem in Fig.~\ref{fig:msg-size-4KB} and Fig.~\ref{fig:qp-num}. 
%\jz{Connect to the Sec.3.2 to explain why}%(more QPs means linearly grabbing more fair shares), which is known to be fair when data sizes are the same but. 
%However, our experiment on the host-FPGA prototype gives the following insights. 

We first run two tenants with the same priority both invoking DMA reads. 
We vary each tenant's DMA read sizes from 16B to 4096B. %(values are 2$^{N}$, N is from 4 to 12) and co-locate them together respectively, i.e., 81 cases in total. 
We simply use robin robin on the FPGA to arbitrate traffic, similar to many other devices~\cite{intel-quickassist}. 
Fig.~\ref{fig:fpga-results}(a) shows representative data points; we omit the others due to space limits. 

\myparagraph{Observation 7: Distinguishing throughput metrics matter}
%\textit{Pitfall 1: Throughput fairness is binary}. 
Intuitively, throughput allocation across two tenants with the same priority can be considered as binary: fair or unfair. 
However, Fig.~\ref{fig:fpga-results}(a) demonstrates that the two tenants can be IOPS fair but data rate unfair (512B + 64B case), or data rate fair but IOPS unfair (1KB + 4KB case). 
Therefore, throughput fairness should be ternary (including neither metric is fair). 
Extending from fairness to arbitrary SLAs of each tenant, providers must distinguish IOPS SLA and data rate SLA to guarantee the required weighted fair share. % of end-to-end I/o accelerator throughput. 
%Different throughput metrics require different contention management to satisfy the weighted fair share required by an arbitrary I/O stream. 
%For example, we should consider the differences of message size pairs, as shown in Fig.~\ref{fig:fpga-results}(a). 

%\textit{Pitfall 2: Robin robin cannot realize isolation for differently-sized message pairs}.
%We first run two tenants both invoking DMA reads, while two tenants have the same priority. 
%We vary each tenant's message sizes from 16B to 4096B (values are 2$^{N}$, N is from 4 to 12) and co-locate them together respectively, i.e., 81 cases in total. 
%We use simple robin robin on the FPGA to arbitrate traffic from the FPGA to the host. 
%We find 60\% of the cases achieve equal throughput allocation across two tenants, 95\% of which exhibit different message sizes across the two tenants. 
%Based on our experiments, equal throughput allocation can be true not only when both tenants have the same message size, but isolation can also be achieved for different message sizes. 
%This is counter-intuitive to datacenter networks where robin robin delivers unfairness for differently-sized message pairs. 

\begin{table}
\centering
\small
\caption{Host-FPGA contention (both tenants send DMA reads). We vary $MaxReadReqSize$ as the PCIe MTU size variations.  % Round robin on FPGA
%Weighted fair share satisfaction table, with the relation between message sizes and PCIe MTU sizes across two tenants. A round robin policy is on FPGA. The two tenants run DMA reads. MTU of PCIe reads here refers to $MaxReadReqSize$.  %\jz{DMA writes define tenants in a different way. }
}
%\vspace{-1em}
\begin{tabular}{c|c|cc} 
\hline
\multicolumn{1}{c}{~} & \multicolumn{1}{c}{} & \multicolumn{2}{c}{Tenant 1 message size M1 (Bytes)}  \\ 
\cline{3-4}
\multicolumn{1}{c}{}  & \multicolumn{1}{c}{} & M1 $<$= MTU    & M1 $>$ MTU                                 \\ 
\hline
Tenant2~              & M2 $<$=                 & IOPS fair        & unpredictable                                  \\
message               & MTU                  & Gbps unfair       & contention                              \\ 
\cline{2-4}
size M2               & M2 $>$                  & unpredictable      & IOPS unfair                                   \\
(Bytes)               & MTU                  & ~contention & Gbps fair                               \\
\hline
\end{tabular}

\label{table:fpga-mtu-msgsize}
\end{table}

\myparagraph{Observation 8: Scheduling of mixed traffic on PCIe interconnects matters} 
IOPS fairness occurs when both tenants send DMA reads whose sizes are smaller than PCIe's $MaxReadReqSize$. 
Data rate (Gbps) is fair when both send messages larger than $MaxReadReqSize$ of PCIe. 
Otherwise, unpredictable contention will occur as summarized in Table~\ref{table:fpga-mtu-msgsize}.
%The root cause is that PCIe interconnect sends one single TLP packet if a message is smaller than $MaxReadReqSize$. 
%However, needs to split it into multiple TLP packets, taking multiple 
We perform the same experiment for two tenants running DMA writes. 
Results are similar to Table~\ref{table:fpga-mtu-msgsize}. 
The only difference is that the boundary is not $MaxReadReqSize$ of PCIe but $MaxPayloadSize$ of PCIe. 
Therefore, being aware of DMA read and write sizes or even re-sizing DMA messages to be either all larger or all smaller than PCIe MTU sizes is important.
By doing so, providers can better guarantee predictable performance for either IOPS or data rate SLAs. 
%In addition, improving fair queuing policies beyond round robin should take such observations into account. 

%\myparagraph{Observation 10: Device-side message size re-splitting helps}
%Mixed message sizes lead to unpredictable contention.
%We find that if the DMA engine has the ability of re-splitting message sizes for each tenant, there should be a case where the contention is much more predictable, if not eliminable. 
%This message size re-sizing should also be aware of PCIe ``MTU" size, in order to not suffering from the contention in Fig.~\ref{}. 

\myparagraph{Observation 9: Colocating opposite directions improve DMA performance predictability}
Given the contention Observation 5, one may consider avoiding colocating DMA writes with DMA reads.
To explore if the DMA read-write mixture is the reason for contention, we run two cases. 
Case 1 is where tenant A sends 4KB DMA reads, and tenant B sends DMA writes, varying its size from 256B to 4KB. 
Case 2 is where tenant A sends 64B DMA reads, and tenant B sends 256B--4KB DMA reads. 
Fig.~\ref{fig:fpga-results}(b) shows that in both cases, tenant B's DMA read performance has a near-zero variance, while tenant A's DMA write performance only varies within 10\%. 
The root cause of this predictability benefit is that DMA reads and writes naturally take advantage of the full-duplex feature of PCIe interconnects. 
Moreover, re-scheduling the timing of DMA reads and writes, and colocating DMA read-intensive and DMA write-intensive traffic patterns together, can improve performance predictability.

%https://ilias.giechaskiel.com/papers/2021_1_cartography_fccm.pdf

%This also implies that the contention Fig.~\ref{fig:dma-direction} is not due to DMA read-write mixture. 
%We guess it may be due to some imperfection on Mellanox NIC interface (e.g., cache misses or buffer contention between RDMA reads and writes).

%\begin{itemize}
%    \item Pitfall: RR doesn't necessarily deliver fairness for mixed message sizes
%    \item Pitfall: DMA traffic doesn't always contend (directions matter, message size pairs matter). 
%    \item Pitfall: Small message would not necessarily mess up the host network (unlike Ethernet). 
%    \item Pitfall: Overall throughput < maximized PCIe throughput does not mean no contention occurs. 
%\end{itemize}

\myparagraph{Observation 10: Preventing small DMA messages from taking over the PCIe interconnects is vital}
Fig.~\ref{fig:fpga-results}(c) and Fig.~\ref{fig:fpga-results}(d) show that the overall PCIe throughput can significantly drop due to small DMA messages.
Cases 3-5 co-locate 4KB DMA writes with $X$ B DMA writes, where $X$ is 64B, 32B, and 16B. 
In Fig.~\ref{fig:fpga-results}(c), Case 3 delivers 51Gbps overall throughput, 94\% of the ideal throughput for DMA writes in our prototype.  
However, Cases 4 and 5 suffer from 28\% and 85\% throughput drops because 32B or 16B DMA writes start to take over PCIe credits within the root complex. 
In Fig.~\ref{fig:fpga-results}(d), $Y$ is 256B, 128B, and 32B, and similar throughput drops are observable. 
Disallowing users to inject such traffic is vital to prevent SLA violations for victim tenants. 
Such small message traffic can be flattened by batching, padding, or rate limiting.

%PCIe lacks sufficient service-level isolation support in the hardware. 
%Although it is constructed with physical lanes that provide natural isolation,
%\footnote{PCIe 3.0 with $\sim$1GB/s per lane, and later generations linearly scale up per-lane bandwidth.}, 
%PCIe bandwidth is not allocated by \textit{lanes} but by \textit{credits}, due to credit-based flow control used in standard PCIe TLP. 
% Reference: https://community.intel.com/t5/FPGA-Wiki/How-Credit-Works-In-Stratix-V-PCIe-G3x1-Reference-Design/ta-p/735609
% Reference: https://www.intel.com/content/www/us/en/docs/programmable/683660/15-1/throughput-optimization.html
%Each credit can potentially leverage more than one lane, and different credits can share the same lane at a given time. 
%Additionally, prior studies show that PCIe bandwidth grows non-linearly as the message size~\cite{neugebauer2018understandingpcie}.
%Therefore, even if credits are exclusively owned by a VM, PCIe contention still exists.

\section{Accelerator Traffic Shaping}\label{sec:design}

\begin{figure}
  \centering
  \includegraphics[width=0.99\columnwidth]{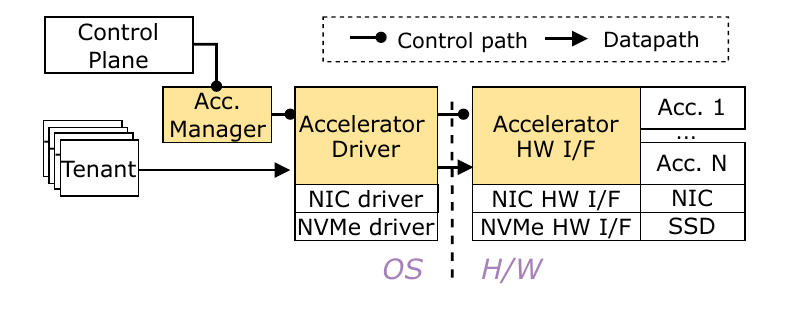}
  %\vspace{-1.5em}
  \caption{A system w/ accelerator management stack. } 
  %\vspace{-1em}
     \label{fig:system}
\end{figure}

Based on observations in Sec.~\ref{sec:understanding}, we advocate one promising approach, i.e., proactively shaping the traffic on-the-fly and highlight the following design options.

\noindent\textbf{Should we support separate accelerator QPs?} 
Existing NICs invoke their accelerators by coupling them with the NIC's QPs~\cite{eran2019nica,eran2022flexdriver,lin2020panic,dpdk-compression-accelerator,dpdk-qat-compression}. 
%Reusing NIC QPs means that accelerators are hidden from the control plane and dataplane applications. 
%In this way, the accelerator can be tightly coupled to a particular network I/O path.  
%This is feasible to incorporate accelerators in inline mode into network applications without software modifications (e.g., drivers). 
However, once the accelerator is allocated, this static provisioning lacks sufficient support to reallocate its compute capacity, and experiences on-NIC contention induced by network traffic. 
% rigidly
As shown in Fig.~\ref{fig:system}, we propose a separate set of accelerator QPs from the host perspective. 
%Fig.~\ref{fig:system} shows a system overview that has an accelerator interface in hardware and separate accelerator driver in the OS. 
%Thus, the OS takes charge of the accelerator resource management with a separate management stack. 
%The stack is in parallel with the NIC and NVMe SSD management stacks, rather than being coupled within them. 
With accelerator QPs, accelerators have a traffic management interface that brings information through customized descriptors.
Designers can optimize the accelerator driver for performance~\cite{wang2020fpgacloud,pirelli2020fasternicdriver} and enable other commands such as initialization, configuration, (re)allocation, monitoring, and traffic shaping. % from the accelerator manager to the accelerator interface. \nej{This sentence is very long and hard to parse}
%Therefore, the manageability for each accelerator I/O flow is improved. 
For example, the provider can control how many QPs are (re)allocated to each flow for performance or contention management (as observed in Sec.~\ref{sec:understanding}).  
%We also articulate that this system design offers a traffic shaping abstraction for us to manage accelerator resources. 
%\jz{This paper~\cite{wang2020fpgacloud} studies the driver efficiency's impact and can be taken care of when we designing our accelerator drivers (also study from NIC drivers~\cite{dpdk-poll-mode-driver} and NVMe drivers~\cite{yang2018spdknvme}.  }

%\mg{The above part talks about QP allocation. How is this related to the key topic of traffic shaping? This needs some explanation.}

\noindent\textbf{Where should the accelerator traffic shaping functionality reside?}
%\kl{You may want to specify what exactly these locations mean}
There are several options for the location of the traffic shaping: software hypervisor on the host, user domain (in VMs or containers), or the accelerator interface on the device. 
First, hypervisors do not always have permission to know detailed traffic information such as message sizes. 
%Additionally, due to host resource contention, high accuracy traffic shaping is sometimes challenging when the hypervisors run in host software.\footnote{For example, to rate limit 100Gbps stably, the software needs to pace 1KB accelerator I/O message every $\sim$80ns on average. Even high-resolution timers in today's software cannot always guarantee such accuracy. } 
Worse, %even though in some cases the inaccuracy is acceptable, 
in-hypervisor shaping loses manageability in some systems where hypervisors are bypassed~\cite{sr-iov,kwon2020fvm,azuresmartnic}. 
Second, users know their own traffic patterns, 
but they do not know the accelerators' characteristics and patterns of other tenants. 
%themselves  have detailed knowledge to shape their traffic patterns, but they may not know the characteristics of the accelerator hosted by the provider. 
%In addition, providers cannot rely on users to shape themselves because users are untrusted in the public cloud. 

Our work advocates traffic shaping in the accelerator interface.  
This design option stands out due to the following reasons. 
First, the accelerator interface has local accessibility to the accelerators. 
Thus, it has on-the-fly traffic observability to accelerators' status, such as queuing effects and egress/ingress bandwidth requirements. 
Second, the interface sits in the middle of diverse accelerator I/O paths, providing good interposition. 
%Thus, it can either work as a switch to forward accelerator traffic in a certain pattern, or perform on-path processing such as metadata interpretation, scatter-gather, address translation, host network traffic management, and telemetry with hardware counters.  
Third, the interface is customizable by providers, e.g., the descriptors for host-accelerator protocol. %One of the successful feature customized on the device is Azure network virtualization~\cite{azuresmartnic}. 

\noindent\textbf{What traffic metrics to shape?}
The traffic shaping unit for each tenant manages the following parameters: message sizes, burst size of messages, the number of QPs to send/receive, and the minimum rate\footnote{The rate of an accelerator traffic stream can be Gbps or IOPS. } (i.e., the SLA), the maximum rate (i.e., how much maximal performance is provisioned). 
Note that because of the non-linearity characteristics of accelerators (Sec.~\ref{characterization:accelerators}), the traffic shaping parameters should be re-calculated case-by-case to match users' requirements under a particular traffic pattern. 

\begin{figure}
  \centering
  \includegraphics[width=0.98\columnwidth]{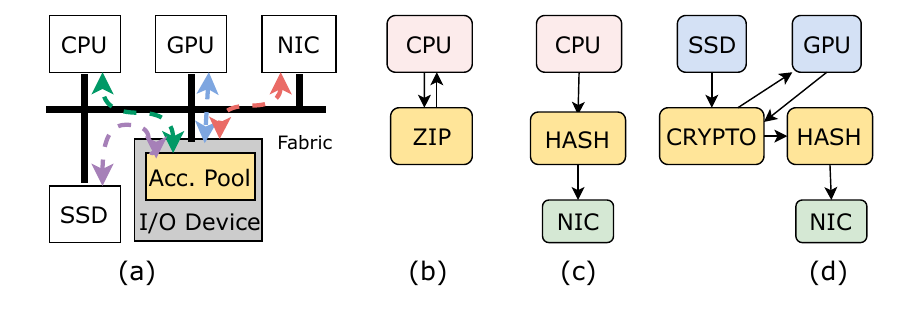}
  %\vspace{-1.5em}
  \caption{End-to-end scenarios. (a) A typical heterogeneous server with wild I/O contention, (b) function call mode, %\footnote{User application running on CPUs calls an I/O accelerator as a function, whose return value goes back to the application (Fig.~\ref{fig:background}(b)). } 
  (c) inline mode, %\footnote{Data is processed by the accelerator while being transferred between the host and the device or between two devices, without CPU involvement for the accelerator invocations  (Fig.~\ref{fig:background}(c)-(d)). } 
  and (d) a complex use case. 
  %\nej{in (a), the NIC is separate from the accelerators but then it appears that in the other figs, the NIC is together with the accelerators?}\jz{If you mean (b)(c)(d), I want to distinguish the NIC and the Net here. I showed Net because Net is the link concept (e.g. Ethernet) but the Net block does not necessarily mean the NIC? Does that make sense? }
  %\nej{I'm not a huge fan of how this figure is drawn -- you have lines connecting Zip to the Acc. Pool from (a) but the figures are quite different overall.}\jz{I see. I've slightly modified the figs.}
  %\nej{I'm not convinced that this figure adds a lot} \jz{do not mention the modes. Overclaim?}\jz{Maybe I can move it to Sec.5, open problems?}
  }
  %\vspace{-1em}
  \label{fig:background}
\end{figure}

\section{Open Problems}\label{sec:open}

%\jz{Add connections between Sec.4 and Sec.5}

\myparagraph{Managing I/O contention in the wild}
A typical server can be like Fig.~\ref{fig:background}(a). 
In such a system, wild I/O contention can occur beyond just accelerator I/O flows. 
For example, an Azure FPGA or an Nvidia BlueField may incorporate both accelerator and network I/O traffic to the host, potentially contending on host networks (similar to Sec.~\ref{characterization:commoditynic}--Sec.~\ref{characterization:fpga}).
%The major difference between shaping accelerator I/O flows and other legacy I/O flows is: besides PCIe interconnects characteristics observed in Sec.~\ref{characterization:commoditynic}--Sec.~\ref{characterization:fpga}, accelerator I/Os should further couple computational characteristics on top of the PCIe traffic shaping. 
Fortunately, our observations generally help improve multi-tenancy for arbitrary I/Os. 
To shape all PCIe traffic, future work can exploit a hybrid traffic shaping approach. 
In a PCIe network, the per-device interface and the PCIe switch can synergistically police outgoing traffic. % by techniques such as priority queues and token buckets or new algorithms that fit PCIe networks. 
This is similar to hybrid traffic shaping in datacenter networks coordinating in-switch and per end-host traffic shaping~\cite{radhakrishnan2014senic,ghobadi2012trickle,saeed2017carousel}. 
When observing traffic patterns on-the-fly, harmful traffic patterns can be used as prior knowledge. 
Researchers can explore various shaping algorithms such as batching, splitting, padding, delaying, interleaving, and rate limiting for performance predictability. 
%\jz{Record a traffic list}

\myparagraph{Assisting congestion control designs}
Intra-host congestion control has become an important topic recently for network applications~\cite{agarwal2023hostcc,liu2023hostping}. 
Several of these works emphasize IOMMU and memory bandwidth related congestion~\cite{agarwal2022understandinghostinterconnect,agarwal2023hostcc}.
Our observations give another dimension of information to reduce congestion if re-shaping the traffic right.  
Our experiments disable IOMMU and isolate NUMA nodes (Sec.~\ref{characterization:method}), orthogonal to findings in prior works~\cite{agarwal2022understandinghostinterconnect,agarwal2023hostcc}. Future work may have further findings when combining our settings with those prior works. 
In addition, our approach will have broader manageability beyond network application, e.g., understanding congestion better when computational I/Os and legacy I/Os co-exist, under a certain PCIe topology. 
%Congestion control techniques from networks~\cite{mittal2015timely} can also be re-designed in our context (lossless PCIe networks). 
%\nej{I don't really know what you mean by demystify IOMMU and memory bandwidth contention?} 

\myparagraph{End-to-end accelerator integration}
There remains discussion of how end-to-end I/O paths can integrate accelerators as Fig.~\ref{fig:background}(b)(c) show. 
Accelerator QPs should exchange accelerator descriptors with other device QPs that exhibit different protocols and descriptor formats. 
One option is to leverage the OS to perform protocol conversion. 
One example is NVMe-over-Fabric protocol~\cite{nvme-of}. 
The other option learns from Internet routers to construct an overlay among diverse QPs across different devices. 
One example of such an IP layer like overlay~\cite{shu2019dua} eliminates host involvements if designed on a PCIe switch or on the device.  

\myparagraph{Accelerator chaining} 
Various aspects of the server’s traffic need to chain different types of services together~\cite{de2011application,katsikas2018metron,zhang2018gnet}.
For instance, in Fig.~\ref{fig:background}(d), after data is read from SSDs, a decryption accelerator is used before the data is fed into the GPU, whose output result is first encrypted and then SHA'ed before sending data into the network. 
Our approach is extensible to accelerator chaining by only shaping ingress and egress patterns of a chain.
Further challenges emerge such as how to manage on-chain traffic and resources, and how to invoke network-attached accelerators if a single device cannot satisfy the entire accelerator chain. 

\myparagraph{Managing I/O contention for GPUs}
%A discrete accelerator unit is usually under time multiplexing without spatial sharing. 
%We can extend our approach to manage accelerators with spatial architectures, e.g., processing element (PE) arrays and GPUs. 
%Our design is also feasible to GPU devices if they are equipped with separate QPs and traffic shaping units between the spatial SIMD stream multiprocessors and the PCIe fabric. 
In a time multiplexing mode, e.g., serving model inferences one user after another, the GPU is equivalent to a temporally multiplexed I/O accelerator. 
In that case, our design is feasible to shape GPU traffic with a PCIe network. 
Some of the big open problems when applying our design to this setting will be \mycircle{1} how to perform traffic shaping when GPUs are used under spatial multiplexing, \mycircle{2} how to incorporate the understanding of GPU internal contention into the traffic patterns to re-shape, and \mycircle{3} how to incorporate our findings when managing PCIe congestion for multi-GPU servers.
%\nej{this open problem feels like it comes out of the blue without a lot of connection to the rest of the paper. }
%\jz{How the contention combinationally work with intra-host congestion control is another problem}
%\jz{Some works target at intra-host PCIe congestion for NICs (SIGCOMM23) or GPUs (HPCA'20, ETH HPC)}
%\jz{GPUs need message sizes at several MBs. Our methodology can benefit their multi-tenancy studies. }

\myparagraph{Fitting future fabrics}
New fabrics like \verb|CXL.io| rely on PCIe interconnects, therefore the observations in Sec.~\ref{sec:understanding} can also guide contention management for CXL devices.
However, CXL devices~\cite{schuh2024ccnic,novakovic2014sonuma} may further bring in new traffic characteristics that either can be leveraged or should be avoided. 
In the future, researchers can exploit them based on methodologies in Sec.~\ref{sec:understanding}, and design appropriate shaping algorithms. 

\myparagraph{Security issues of accelerator sharing}
We can quantify a set of attacks, such as performance attacks due to small messages (Observation 9). 
Further study of the security implications of accelerator sharing is needed. 
For example, newly-introduced side channels can be a new vulnerability if users leverage cryptography accelerators to accelerate security operations; the sharing degree between providers and users remains another question to explore. 
%\jz{Contention management can be the foundation of Multi-party computation between users that program the accelerator devices, third-party device vendors and cloud providers. }

\myparagraph{Accelerator cost and cloud-scale management}
Generally, sharing reduces accelerator cost, e.g., TCO or embodied carbon emissions~\cite{sadok2023apples,plotnik2024intergenerational}. 
Research on holistic cost models will help business profits and sustainability.
For example, using densely populated accelerator servers~\cite{lim2024beehive,taylor2020asiccloud,tpuv4} or attaching homogeneous accelerator devices per server~\cite{aws-nitro,azuresmartnic} is an open question. 
In addition, this paper only studies the case of two-tenant co-location. 
With more tenants, there should be new capacity planning and admission control supports at cloud-scale, balancing costs and serviceability. 

\bibliographystyle{ACM-Reference-Format} 
\bibliography{hotnets24-template}

\end{document}